\documentclass[aps,prb,twocolumn,superscriptaddress]{revtex4-2}
\usepackage{graphicx}
\usepackage{physics}
\usepackage{advdate}
\usepackage{comment}
\usepackage[colorlinks=true]{hyperref}

\newcommand{\figref}[1]{\figurename~\ref{#1}}

\usepackage{xcolor}
\usepackage{soul}
\definecolor{honey}{HTML}{e29006}
\definecolor{honey-light}{HTML}{f8e8d2}

\usepackage[normalem]{ulem}

\newcommand{\omsp}{\omega _ \mathrm{s}}

\begin{document}

\title{Quantum Friction near the Instability Threshold}

\author{Daigo Oue}
\email{daigo.oue@gmail.com}
\affiliation{
  Instituto Superior T\'{e}cnico--University of Lisbon and Instituto de Telecomunica\c{c}\~{o}es, Lisbon 1049-001, Portugal
}
\affiliation{
    Innovative Photon Manipulation Research Team, RIKEN Center for Advanced Photonics, Saitama 351-0198, Japan
}
\affiliation{
    The Blackett Laboratory, Imperial College London, London SW7 2AZ, United Kingdom
}

\author{Boris Shapiro}
\email{boris@physics.technion.ac.il}
\affiliation{Department of Physics, Technion--Israel Institute of Technology, Haifa 32000, Israel}

\author{M\'{a}rio G. Silveirinha}
\email{mario.silveirinha@tecnico.ulisboa.pt}
\affiliation{
  Instituto Superior T\'{e}cnico--University of Lisbon and Instituto de Telecomunica\c{c}\~{o}es, Lisbon 1049-001, Portugal
}

\date{\today}

\begin{abstract}
  In this work, we develop an analytical framework to understand quantum friction across distinct stability regimes, providing approximate expressions for frictional forces both in the deep stable regime and near the critical threshold of instability. Our primary finding is analytical proof that, near the instability threshold, the quantum friction force diverges logarithmically. This result, verified through numerical simulations, sheds light on the behavior of frictional instabilities as the system approaches criticality.
  Our findings offer new insights into the role of instabilities, critical divergence and temperature in frictional dynamics across quantum and classical regimes.
\end{abstract}

\maketitle

\section{Introduction}	

The Casimir effect, a fundamental consequence of quantum fluctuations, generates an attractive force between neutral bodies in vacuum.
This phenomenon was first theoretically predicted by Casimir and Polder in their work on atom-on-a-plate interactions~\cite{casimir1948influence}, now known as the Casimir-Polder effect.
They demonstrated that quantum fluctuations of the electromagnetic field give rise to an attractive force between a neutral atom and a conducting plate.
Extending this idea, Casimir later studied a cavity-type configuration consisting of two neutral, perfectly conducting plates, showing that a similar attractive force occurs between the plates~\cite{casimir1948attraction}.
These theoretical discoveries laid the foundation for a broad range of research into fluctuation-induced phenomena.

While the original Casimir effect deals with two parallel plates in close proximity, the force can be modified by applying spatial modulations to the plates~\cite{chan2008measurement,bao2010casimir,intravaia2013strong,wang2021strong,somers2018measurement,spreng2022recent,kuccukoz2024quantum}.
For instance, nanostructuring the surface of the plates has been shown to alter the magnitude of the Casimir force, offering new methods for controlling and applying this interaction~\cite{chan2008measurement,bao2010casimir,intravaia2013strong,wang2021strong}.
Furthermore, by introducing anisotropy, fluctuation-induced torques can be exerted in addition to the attractive force~\cite{somers2018measurement,spreng2022recent,kuccukoz2024quantum}.
These spatial modulations have expanded the possibilities for utilising the Casimir effect in practical systems.
However, beyond spatial modulations, temporal modulations of the system have also led to new and intriguing discoveries.

One particularly interesting temporal modulation is seen in the dynamical Casimir effect, where real photons are produced from quantum fluctuations due to the periodic motion of one of the plates~\cite{moore1970quantum,fulling1976radiation,wilson2011observation}.
In this scenario, the vertical position of one plate oscillates, causing the separation between the plates to vary periodically.
As a result, quantum fluctuations in the vacuum are converted into real photons.
A related phenomenon, quantum friction, arises when the plates move laterally in opposite directions~\cite{volokitin2008theory,dedkov2017fluctuation,milton2016reality,reiche2022wading,hoye1992friction,brevik1993friction,pendry1997shearing,pendry1998can,volokitin1999theory}.
This lateral motion leads to a dynamic modulation of the quantum vacuum, producing real photons and, consequently, generating a frictional force between the plates~\cite{pendry1998can}.
This frictional force occurs without direct contact between the plates, driven purely by the shearing motion and quantum fluctuations.

Early theoretical studies implicitly assumed that these systems would reach a nonequilibrium steady state, resulting in a time-independent frictional force~\cite{volokitin2008theory,dedkov2017fluctuation,milton2016reality,reiche2022wading,hoye1992friction,brevik1993friction,pendry1997shearing,pendry1998can,volokitin1999theory}. However, recent researches have revealed that quantum frictional systems may experience instabilities under certain conditions~\cite{Friction3, Friction2, Friction1, guo2014singular,brevik2022fluctuational,oue2024stable}. For example,  Brevik et al.~\cite{brevik2022fluctuational} recently employed a nonpertubative approach to demonstrate that such systems can exhibit an unstable behavior where the excitation population grows exponentially over time.
This instability leads to a continuously increasing frictional force, demonstrating that the system does not reach a stable, steady state in some regimes.
Although several previous studies addressed the gap between the two distinct regimes~\cite{guo2014giant,guo2014singular,oue2024stable}, they largely rely on numerics and leave room for further analytical exploration, particularly near the threshold of instability.

In this work, we provide approximate analytical expressions for quantum friction in both the deep stable regime and near the critical regime where the system approaches instability.
The approximate formulae clarify the behavior of the system in these distinct regimes.
Additionally, we generalize the study of friction near the instability threshold to incorporate the effects of thermal fluctuations. This allows us to investigate how the frictional force crosses over from quantum (zero-temperature) to classical (high-temperature) regimes in the vicinity of criticality.
Our results contribute to a deeper understanding of quantum friction and its stability across different regimes.

\section{Instability threshold}
\label{sec:instability}
\begin{figure}[htbp]
  \centering
  \includegraphics[width=\linewidth]{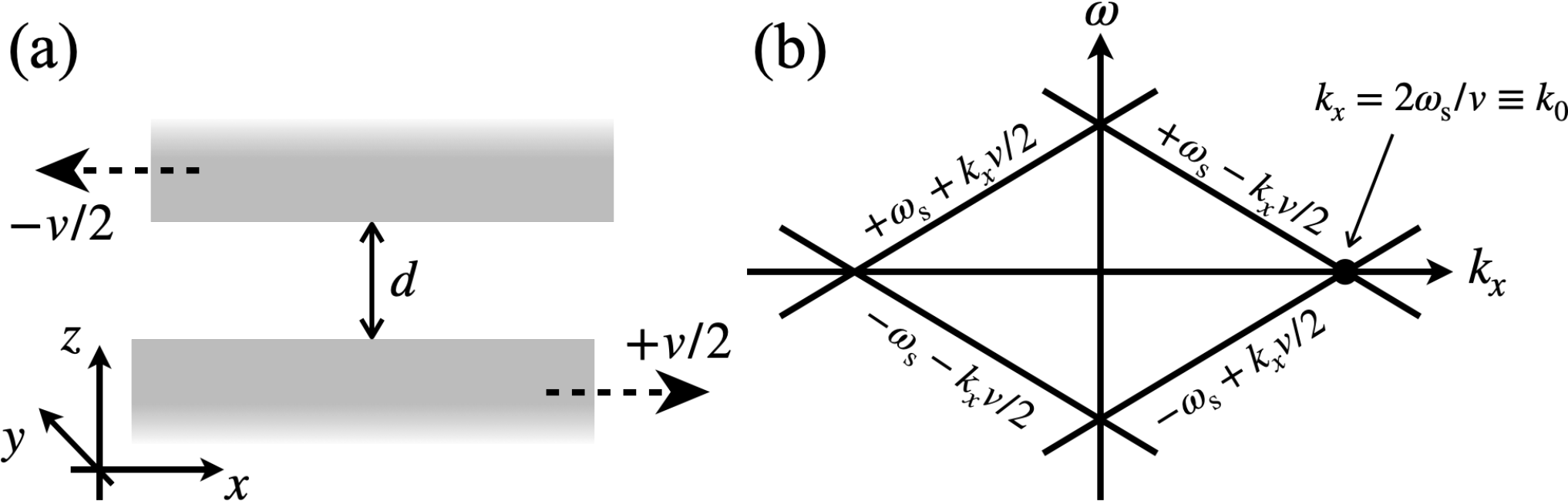}
  \caption{
    (a) The setup studied in this article.
    Two metallic or semiconducting plates are separated by a vacuum gap of width $d$ and in relative motion with speed $v$.
    (b) Dispersion curves without the interaction between the two surfaces.
  }
  \label{fig:setup}
\end{figure}

Consider two metallic or semiconducting plates at a distance $d$ in relative motion [\figref{fig:setup}(a)].
Isolated plates ($d \rightarrow \infty$) can support Doppler-shifted surface plasmons.
For a finite distance, they are hybridized, leading to the dispersion relation~\cite{brevik2022fluctuational,oue2024stable},
\begin{align}
  \omega 
  = \pm \omega _ {\mathrm{a,b}} ^ {(0)}
  = \pm \omsp \sqrt{
    1 + \qty(\frac{k _ x v}{2\omsp}) ^ 2 
    \pm \sqrt{e ^ {-2\abs{\vb{k}}d} + \qty(\frac{k _ x v}{\omsp}) ^ 2 }
  }
  \label{eq:dispersion}
\end{align}
where the subscript ``a (b)'' specifies the sign $+$ ($-$) inside the radical, $\omsp$ is the surface plasmon resonant frequency for an isolated plate, and $\vb{k} = (k _ x, k _ y)$ is the transverse wavevector.
The dispersion equation supports four branches, which include both the positive frequency and negative frequency branches, linked by particle-hole symmetry, as usual (first $\pm$ sign in Eq.~\eqref{eq:dispersion}).
Thus, two independent waves can propagate in the system for a given $\vb{k}$.
The $\pm$ sign inside the square root defines the two distinct branches of the spectrum.
In \figref{fig:setup}(b), we show the dispersion curves in the non-interacting limit ($d \rightarrow \infty$).
Of particular interest are the crossing points, $k _ x = \pm 2\omsp/v \equiv \pm k _ 0$.
For $\vb{k} = (k _ 0, 0)$, the coupling $e ^ {-2k _ 0 d}$, albeit exponentially small, leads to a profound change in the dispersion relation, resulting in purely imaginary frequencies, $\pm \omega _ \mathrm{b} ^ {(0)} \approx \pm i(\omsp/2) e ^ {-k _ 0 d}$.
The root with a positive imaginary part represents an instability (i.e., exponentially growing solution in time).
The instability is not limited to the wave vector $\vb{k} = (k _ 0, 0)$ but occurs in a narrow region of $\vb{k}$ centered around this value~\cite{brevik2022fluctuational, Friction3},
\begin{align}
  1 - \frac{e ^ {-\abs{\vb{k}}d}}{2} < \abs{\frac{k _ x}{k _ 0}} < 1 + \frac{e ^ {-\abs{\vb{k}}d}}{2}.
  \label{eq:instability-window}
\end{align}
Within this region, the growth rate (the positive imaginary part of the corresponding root) is evaluated as~\cite{brevik2022fluctuational, Friction3},
\begin{align} \label{omegaplus}
  \Im \qty{\omega _ \mathrm{b} ^ {(0)}} = \frac{1}{2}\omsp e ^ {-\abs{\vb{k}}d}\sqrt{1 - \qty(\frac{2 - \eta \abs{k _ x}d}{e ^ {-\abs{\vb{k}}d}}) ^ 2},
\end{align}
where $\eta \equiv v/(\omsp d) = 2/(k _ 0 d)$ is the destabilizing factor.
Since quantum (and thermal) fluctuations with all possible $\vb{k}$ exist, the system is unstable even for a small relative velocity $v$.

So far, the discussion neglects the effect of dissipation in the plates. It is characterized by the collision frequency $\gamma$ and modifies the dispersion relation as follows (a quasi-static approximation is implicit, see Ref. \cite{silveirinha2014optical}, Eq. (30)):
\begin{widetext}
\begin{align}
  \omega = -i\omsp
  \qty[
  \frac{\gamma}{2\omsp}
  \pm \sqrt{
    1 + \qty(\frac{k _ x}{k _ 0}) ^ 2 - \qty(\frac{\gamma}{2\omsp}) ^ 2 
    \pm \sqrt{
      e ^ {-2\abs{\vb{k}}d} +
      \qty(\frac{2k _ x}{k _ 0}) ^ 2 
      \qty(1-\frac{\gamma ^ 2}{4\omsp ^ 2})
    }
  }
  ].
  \label{eq:dispersion-modified}
\end{align}
\end{widetext}
The dissipation is a stabilizing factor, allowing for a non-equilibrium stationary state (NESS). Here, we disregard the effect of frictional force on the velocity of the plates, as it is entirely negligible on the timescale relevant to this problem.
Such an NESS exists only if the velocity $v$ is not too large for a given $\gamma$, ensuring that dissipation mechanisms dominate over gain mechanisms \cite{oue2024stable}.  Otherwise, instabilities arise.

To determine the instability threshold, we find the most favorable value of $\vb{k}$ for the instability to occur.
Thus, we set $k _ y = 0$ and write Eq.~\eqref{eq:dispersion-modified} for the potentially ``unstable root,''
\begin{align}
  &\omega _ \mathrm{b} = 
  \notag \\
  &i\omsp\qty[-\Gamma + \sqrt{\sqrt{e ^ {-4\abs{\beta}/\eta} + 4\beta ^ 2\qty(1 - \Gamma ^ 2)} - \qty(1 + \beta ^ 2 - \Gamma ^ 2)}],
\end{align}
where dimensionless parameters, $\Gamma = \gamma/(2\omsp)$ and $\beta = k _ x /k _ 0$, have been introduced.
The system is stable if $\Im \qty{\omega _ \mathrm{b}} < 0$ for all $k _ x$.
This condition is fulfilled if
\begin{align}
  e ^ {-4\abs{\beta}/\eta} - 4\beta ^ 2 \Gamma ^ 2 < (1 - \beta ^ 2) ^ 2
  \qquad
  (\textrm{for all $\beta$}).
  \label{eq:stability-condition}
\end{align}
\underline{(1) Weak dissipation ($\Gamma \ll 1$)}.
To maintain the system stability, it is enough to satisfy Eq.~\eqref{eq:stability-condition} with $\beta = 1$, which is the optimal value for developing an instability in the absence of dissipation~\cite{brevik2022fluctuational, Friction3}.
This yields the instability threshold,
\begin{align}
  e ^ {-4/\eta} = 4\Gamma ^ 2
  \Rightarrow
  \eta = \frac{2}{\log[1/(2\Gamma)]}\equiv \eta _ \mathrm{c}(\Gamma).
  \label{eq:threshold}
\end{align}
Namely, for a fixed, small $\Gamma$, the system is stable if $\eta < \eta _ \mathrm{c}(\Gamma)$ or if $v < v _ \mathrm{c}(\Gamma) \equiv 2\omsp d/\log[1/(2\Gamma)]$.
This result was previously obtained in Refs.~\cite{guo2014giant,oue2024stable}.

\noindent
\underline{(2) Strong dissipation ($\Gamma \simeq 1$)}.
Let us first note that for $\Gamma > 1/\sqrt{2}$, the system is stable for all values of $\beta$.
Indeed, for $\Gamma = 1/\sqrt{2}$, Eq.~\eqref{eq:stability-condition} yields $e ^ {-4\abs{\beta}/\eta} < 1 + \beta ^ 4$ which is satisfied for any $\beta$, although barely so for $\beta \rightarrow 0$.
For $\Gamma$ somewhat smaller than $1/\sqrt{2}$ (i.e., $\Gamma = (1-\delta)/\sqrt{2}$ with $\delta \ll 1$) instability becomes possible for sufficiently small $k _ x$.
A straightforward analysis shows that for a fixed, small $\delta$, the instability threshold is $\eta _ \mathrm{c} \simeq \delta ^ {-3/2} \gg 1$.
The system is stable unless the velocity reaches a large value, $v _ \mathrm{c} \simeq \omsp d \delta ^ {-3/2}$.

\begin{figure}[tbp]
  \centering
  \includegraphics[width=.8\linewidth]{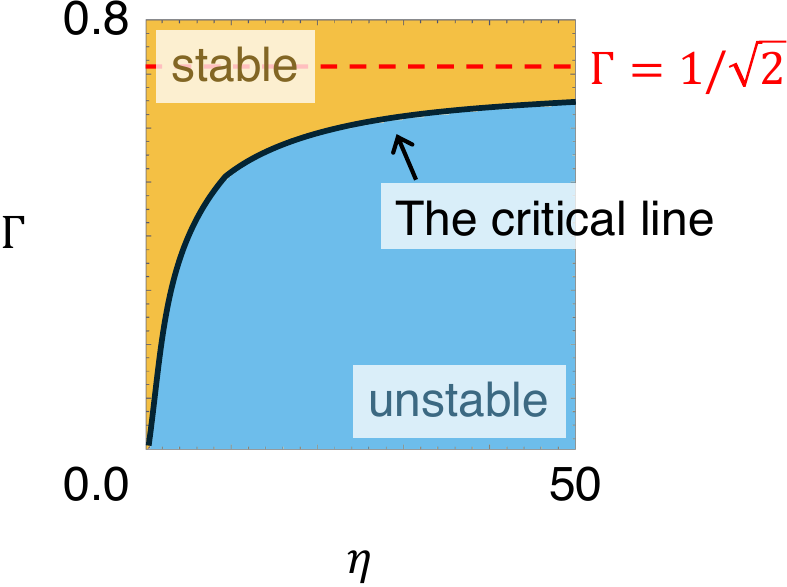}
  \caption{
    Stability--instability diagram.
    The system is always stable ($\Im\qty{\omega _ \mathrm{b}} < 0$) if $\Gamma > 1/\sqrt{2}$.
  }
  \label{fig:diagram}
\end{figure}
The full stability--instability diagram is depicted in \figref{fig:diagram}. It is found by minimizing the function $\Gamma \left( \beta  \right) = {\left( {\frac{{{e^{ - 4\left| \beta  \right|/\eta }} - {{\left( {1 - {\beta ^2}} \right)}^2}}}{{4{\beta ^2}}}} \right)^{1/2}}$ with respect to $\beta$.

\section{Quantum friction and its divergence}

In the stable region (NESS), there is a time-independent friction force $F$ (as previously noted, we neglect changes in the kinetic energy of the plates).
The expression for this force was derived by a number of authors~\cite{volokitin2008theory,dedkov2017fluctuation,oue2024stable}. At zero temperature, the friction force per unit of area is:
\begin{widetext}
\begin{align}
  F = \int \limits _ {-\infty} ^ {+\infty}
  \int \limits _ 0 ^ \infty
  \int \limits _ {-\tfrac{k _ x v}{2}} ^ {+\tfrac{k _ x v}{2}} 
  \frac{\hbar k _ x}{2\pi ^ 3} \frac{\Im\qty{R _ -} \Im\qty{R _ +} e ^ {-2\abs{\vb{k}}d}}{\abs{1 - R _ - R _ + e ^ {-2\abs{\vb{k}}d}} ^ 2}
  \dd{\omega}\dd{k _ x}\dd{k _ y}, 
  \label{eq:F}
\end{align}
\end{widetext}
where 
\begin{align}
  R _ \pm = \frac{\omsp ^ 2}{\omega _ \pm(\omega _ \pm + i\gamma)- \omsp ^ 2},
  \label{eq:R}
\end{align}
are the quasi-static reflection coefficients.
Note that we defined $\omega _ \pm := \omega \pm k _ x v/2$.
The denominator in Eq.~\eqref{eq:F} accounts for the multiple scattering between plates.
Note that Eq.~\eqref{eq:F} holds in the reference frame where both plates are in motion, while the expression in Refs.~\cite{volokitin2008theory,dedkov2017fluctuation,oue2024stable} is in a frame where one of the plates is at rest.
The two expressions are related by the transformation $\omega \mapsto \omega - k _ x v/2$ and are consistent due to the Galilean invariance of $F$.

In the present section, we focus our attention on the weak dissipation regime, $\Gamma \ll 1$, so that the instability threshold for the normalized velocity $\eta$ is given in Eq.~\eqref{eq:threshold}.
Alternatively, by fixing a small value of $\eta$, one can get the instability threshold for $\Gamma$ as $\Gamma _ \mathrm{c} = (1/2)e ^ {-2/\eta} \ll 1$.
We consider two distinct cases in the following.

\noindent
\underline{(i) ``deep stable region'' ($\Gamma _ \mathrm{c} \ll \Gamma \ll 1$)}:
In this case, the system is far away from the instability threshold.
The multiple-scattering effect can then be neglected (i.e., the denominator in Eq.~\eqref{eq:F} is set to unity).
Furthermore, since the integral over $k _ x$ in Eq.~\eqref{eq:F} has a sharp cutoff at $k _ x \sim 1/d$, $k _ x v/2$ is small compared to $\omsp$ and so is $\omega$ when evaluating the imaginary part $\Im\qty{R _ \pm}$; hence,
$
\Im \qty{R _ \pm} \approx -\omega _ \pm\gamma/\omsp ^ 2.
$
After the integration over $\omega$ and change of variables, $k _ x d = x$ and $k _ y d = y$, Eq.~\eqref{eq:F} is reduced to
\begin{align}
  F = -\frac{\hbar\omsp\eta ^ 3}{6\pi ^ 3 d ^ 3} \Gamma ^ 2 \iint x  ^ 4 e ^ {-2\sqrt{x ^ 2 + y ^ 2}}\dd{x}\dd{y}
    = -\frac{15\hbar\omsp\eta ^ 3 \Gamma ^ 2}{64\pi ^ 2 d ^ 3}.
\end{align}
Thus, for small velocities ($\eta \ll 1$), the friction force $F$ is proportional to $v ^ 3$ in the deep stable regime, in agreement with Ref. \cite{pendry1997shearing}.

\noindent
\underline{(ii) Near threshold ($\Gamma \simeq \Gamma _ \mathrm{c}$)}:
We set $\Gamma = \Gamma _ \mathrm{c} (1+\varepsilon)$ with $\varepsilon \ll 1$.
The multiple scattering becomes important in this regime, and the denominator in Eq.~\eqref{eq:F} approaches zero for the appropriate values of $\omega$ and $\vb{k}$.
To study this case, we substitute Eq.~\eqref{eq:R} into Eq.~\eqref{eq:F}.
After some algebra, we obtain
\begin{widetext}
\begin{align}
  &F = \frac{\hbar\omsp ^ 4 \gamma ^ 2}{2\pi ^ 3}
  \int \limits _ {-\infty} ^ {+\infty}
  \int \limits _ 0 ^ \infty
  \int \limits _ {-\tfrac{k _ x v}{2}} ^ {+\tfrac{k _ x v}{2}}
  k _ x \frac{(\omega ^ 2 - k _ x ^ 2 v ^ 2/4) e ^ {-2\abs{\vb{k}}d}}{\abs{\Delta} ^ 2}
  \dd{\omega}\dd{k _ x}\dd{k _ y},
  \label{eq:F-critical}
  \\
  &\Delta = \qty(\omega _ +\qty(\omega _ + + i\gamma) - \omsp ^ 2)
  \qty(\omega _ -\qty(\omega _ - + i\gamma) - \omsp ^ 2)
  - \omsp ^ 4 e ^ {-2\abs{\vb{k}}d}.
\end{align}
\end{widetext}
Note that $\Delta(\omega, k _ x, k _ y) = 0$ is the surface plasmons dispersion relation with the solution given in Eq.~\eqref{eq:dispersion-modified}.
For small $\Gamma$, the four roots of $\Delta$ are approximated as
$
\pm\omega _ \mathrm{a,b} \approx -i\omsp\Gamma \pm \omega _ \mathrm{a,b} ^ {(0)},
$
with $\pm\omega _ \mathrm{a,b} ^ {(0)}$ the four roots in the absence of dissipation defined as in Eq. \eqref{eq:dispersion}. 
We are interested in the roots near the real $\omega$ axis.
Such potentially unstable roots provide the largest contribution to $F$.
For $k _ x \approx k _ 0$ and $k _ y \approx 0$, there is a pair of roots close to the origin,
$\Omega _ {\mathrm{b}\pm} \approx -i\omsp\Gamma \pm i[\omega _ \mathrm{b} ^ {(0)}]'' \equiv i\Omega _ {\mathrm{b}\pm}''$,
with $[\omega _ \mathrm{b} ^ {(0)}]''$ given by Eq. \eqref{omegaplus}. Additionally, there is another pair of roots located at $\omega \approx \pm2\omsp$.

The singular behavior of the friction force is determined by the behavior of the integrand for $\omega$ near the origin, corresponding to the nearly unstable root.
For $\omega \approx 0$, $\abs{\Delta} ^ 2$ can be approximated as $16\omsp ^ 4 [\omega ^ 2 + (\Omega _ {\mathrm{b}+}'') ^ 2][\omega ^ 2 + (\Omega _ {\mathrm{b}-}'') ^ 2]$.
Furthermore, in the term $\omega ^ 2 - k _ x ^ 2 v ^ 2/4$ in Eq.~\eqref{eq:F-critical}, $\omega ^ 2$ can be neglected so that
\begin{widetext}
\begin{align}
  F \approx \frac{\hbar \gamma ^ 2}{32\pi ^ 3}
  \int \limits _ {-\infty} ^ {+\infty} \,\,
  \int \limits _ {k_0 - \tfrac{k_0}{2} f(k_y)} ^ {k_0 + \tfrac{k_0}{2} f(k_y)} \,\,
  \int \limits _ {-\tfrac{k _ x v}{2}} ^ {+\tfrac{k _ x v}{2}}
  \frac{-(k _ x ^ 3 v ^ 2/4) e ^ {-2\abs{\vb{k}}d}}{[\omega ^ 2 + (\Omega _ {\mathrm{b}+}'') ^ 2][\omega ^ 2 + (\Omega _ {\mathrm{b}-}'') ^ 2]}
  \dd{\omega}\dd{k _ x}\dd{k _ y},
  \label{eq:F-critical-approx}
\end{align}
\end{widetext}
where we defined $f(k _ y) = e ^ {-\sqrt{k _ 0 ^ 2 + k _ y ^ 2}d}$.
Note that we restricted the integration range in $k_x$ to the narrow window $\left| {{k_x} - {k_0}} \right| < (k_0/2)f(k_y)$ centered at $k _ x = k _ 0$ [Eq.~\eqref{eq:instability-window}].

In the integral over $\omega$, the integration limits can be extended to $\pm \infty$, as the relevant value of $k _ x v/2 \approx k _ 0 v/2 = \omsp$ is comfortably away from the dominant integration region $\omega \approx 0$. With this approximation, the integral over $\omega$ is $\pi/(\gamma \Omega _ {\mathrm{b}+}''\Omega _ {\mathrm{b}-}'')$. Setting all $k _ x$ in the integrand of Eq.~\eqref{eq:F-critical-approx} identical to $k _ 0$ one finds that:
\begin{align}
  F \approx \frac{\hbar \omega_{\rm{s}}^3 \gamma}{16 \pi ^2 v}
  \int \limits _ {-\infty} ^ {+\infty}
  \,\,\int \limits _ {k_0 - \tfrac{k_0}{2} f} ^ {k_0 + \tfrac{k_0}{2} f}
  \frac{-\dd{k _ x}}{\Omega _ {\mathrm{b}+}'' \Omega _ {\mathrm{b}-}''}
  \qty{f(k _ y)} ^ 2\dd{k _ y}.
  \label{eq:F-critical-approx2}
\end{align}
Introducing the variable $q = 2(k _ x /k _ 0 - 1)/f$, we obtain
\begin{align}
  F \approx \frac{\hbar \omsp^2 \gamma}{4\pi ^ 2 v ^ 2}  
  \int \limits _ {-\infty} ^ {+\infty}
  \int \limits _ {-1} ^ {+1}
  \frac{ -f(k _ y) \dd{q}}{4\Gamma ^ 2 - \qty{f(k _ y)} ^ 2 (1-q ^ 2)}\qty{f(k _ y)}^2\dd{k _ y}.
  \label{eq:F-critical-approx-cv}
\end{align}
The integration over $q$ can be done analytically and yields:
\begin{align}
  F \approx -\frac{\hbar \omsp^3}{2\pi ^ 2 v ^ 2}  
  \int \limits _ {-\infty} ^ {+\infty}
  \frac{{{\rm{arcsin}}\qty[ {{\omega _{\rm{s}}}f(k_y)/\gamma } ]}}{{\sqrt {1 - {{\qty[ {{\omega _{\rm{s}}}f(k_y)/\gamma } ]}^2}} }}
  \qty{f(k_y)} ^ 2 \dd{k _ y}.
  \label{eq:F-critical-final}
\end{align}

Next, we set $\Gamma = \Gamma _ \mathrm{c} (1 + \varepsilon) = (1/2)e ^ {-2/\eta}(1 + \varepsilon)$ and concentrate on the term ${({{{\omega _{\rm{s}}}}}/{\gamma }){f(k_y)}}$. It can be expanded in a Taylor series in $k_y$ as follows: $({{{\omega _{\rm{s}}}}}/{\gamma }){f(k_y)} \approx ({{{\Gamma _{{\rm{c}}}}}}/{\Gamma })\left( {1 - {{k_y^2d}}/({{2{k_0}}})} \right)$.
Keeping the dependence on $k_y$ in the denominator is crucial: setting $k _ y = 0$ in the term ${({{{\omega _{\rm{s}}}}}/{\gamma }){f(k_y)}}$ would result in a square root singularity $F \sim 1/\sqrt{\varepsilon}$, while the more accurate treatment smooths out this behavior into a weaker logarithmic divergence, as shown below.

Near the singularity, the $\arcsin$ term in the integrand can be replaced by $\pi/2$. Furthermore, the prefactor in front of the integral measure can be evaluated at $k_y=0$ [i.e., $\qty{f(k_y)} ^ 2\dd{k _ y} \approx \qty{f(0)} ^ 2\dd{k _ y} = e ^ {-4/\eta} \dd{k _ y} $] and moved out of the integrand if the integration range is truncated to $\left| {{k_y}} \right| < {k_{y,\max }}$. These approximations yield:
\begin{align}
  F \approx -\frac{\hbar \omsp^3}{4\pi v ^ 2} {e^{ - 4/ \eta}} 
  \int \limits _ {-k_{y,\max }} ^ {+k_{y,\max }}
  \frac{1}{{\sqrt { {2 \varepsilon + \frac{{k_y^2d}}{{{k_0}}}}  } }} \dd{k _ y}.
  \label{eq:F-critical-final2}
\end{align}
Near the instability threshold ($\varepsilon \to 0^+$), the integral evaluates to $\sqrt {\frac{{{k_0}}}{d}} \log \left( {\frac{{2k_{y,\max }^2d}}{{{k_0}\varepsilon }}} \right)$. We choose $k_{y,\max }$ in such a way that the quadratic term in the Taylor expansion is small as compared to the leading term. Specifically, we take $
\frac{{k_{y,\max }^2 d}}{{2{k_0}}} = \frac{1}{4}$. This final approximation gives the analytical result: 
\begin{align}
  F \approx -\frac{\hbar\omsp}{2\pi\sqrt{2}d ^ 3}\eta ^ {-5/2} e ^ {-4/\eta} \log \frac{1}{\varepsilon}.
  \label{eq:F:log-divergence}
\end{align}
As seen, the friction force diverges logarithmically near the instability threshold, in agreement with previous numerical results \cite{guo2014singular}. The above result is valid under the conditions $\varepsilon \ll 1$ and $e^{-4/\eta} \ll 1$.  We underline that it is essential to keep the $k_y ^ 2$ correction in Eq. \eqref{eq:F-critical-final2} to obtain the logarithmic behavior of $F$. Although our focus in this article is on the behavior of the force as the collision frequency approaches criticality, our result is universal. In particular, a similar logarithmic divergence arises when the relative velocity approaches the critical value at a fixed level of dissipation.

\begin{figure}
    \centering
    \includegraphics[width=\linewidth]{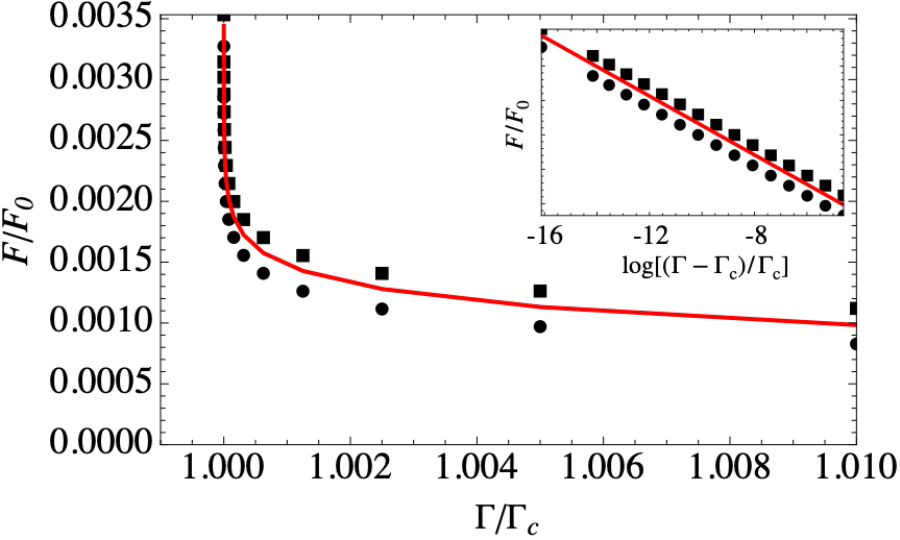}
    \caption{
    Comparison of the approximate formulae for the friction force with the exact result for $\eta = v/(\omsp d) = 0.5$.
    The square markers represent the exact result \eqref{eq:F}, the circle markers the approximate integral \eqref{eq:F-critical-final}, and the red curve the analytical formula \eqref{eq:F:log-divergence}.
    The two approximate formulae agree well with the exact result around the instability threshold.
    The inset shows the normalized force as a function of $\log(\varepsilon)$ and confirms the logarithmic divergence near criticality.
    The force normalization factor is set $F_0 = \hbar\omsp/d ^ 3$.
    }
    \label{fig:comparison}
\end{figure}

In \figref{fig:comparison}, we compare the approximate formulae, Eqs.\,\eqref{eq:F-critical-final} and \eqref{eq:F:log-divergence}, with the exact result \eqref{eq:F}.
It is evident that the approximate formulae work well near the instability threshold.
Furthermore, they accurately predict the logarithmic divergence near criticality, as illustrated in the inset of \figref{fig:comparison}.
For some semiconductors, $\omsp/\gamma$ can be as large as $\omsp/\gamma = 10$ (see, e.g., \cite{palik1976coupled,majorel2019theory}), and the critical point is roughly estimated as $\eta _ \mathrm{c} \sim 1$ (e.g., $v _ \mathrm{c} \sim 10 ^ 5\ \mathrm{m/s}$ for $\omsp/(2\pi) = 1.5\ \mathrm{THz}$ and $d = 10\ \mathrm{nm}$). 
When the instability threshold is approached, the friction force exhibits the logaarithmic behaviour $F \sim 0.002 \times F _ 0 \log (\varepsilon)$, where we defined $F _ 0 = \hbar\omsp/d ^ 3$.
For a plasma frequency $\omsp/(2\pi) \sim 1.5\ \mathrm{THz}$ (e.g., narrow gap semiconductors such as InSb or InP~\cite{palik1976coupled,gomez2005transmission}) and for a separation width $d = 10\ \mathrm{nm}$, the prefactor can be estimated as $0.002 \times F _ 0 \approx 2.0\ \mathrm{pN/\mu m ^ 2}$.

When thermal fluctuations are present alongside quantum fluctuations, the integration limits in the integral over $\omega$ in Eq.~\eqref{eq:F} should be extended to $\pm\infty$. Additionally,  a ``smoothing factor,'' 
\begin{align}
\frac{1}{2}\qty{\coth\frac{\hbar\omega _ +}{2k _ \mathrm{B} T} - \coth\frac{\hbar\omega _ -}{2k _ \mathrm{B} T}}, \nonumber
\end{align}
should be introduced~\cite{dedkov2017fluctuation,footnote1} in the integrand.
The behavior of the friction force near criticality is determined mainly by the singular multi-reflection factor in the integrand of Eq.~\eqref{eq:F}. Thus, the asymptotic formula for the friction force at a finite temperature can be readily obtained by multiplying Eq. \eqref{eq:F:log-divergence} by the smoothing factor evaluated at $k_x=k_0$ and $\omega=0$, which corresponds to the singularity of the integrand. This yields: 
\begin{align}
  F \approx - \frac{\hbar\omsp}{2\pi\sqrt{2}d ^ 3}  \eta ^ {-5/2} e ^ {-4/\eta} \coth\frac{\hbar \omsp }{2k _ \mathrm{B} T} \log \frac{1}{\varepsilon}.
  \label{eq:finiteT}
\end{align}
In particular, in the high-temperature classical limit, $k _ \mathrm{B} T \gg \hbar\omsp$,
the $\hbar$ drops out from the formulas, and the force near the instability threshold is larger by a factor $2 k _ \mathrm{B} T/(\hbar\omsp)$ than its purely quantum counterpart calculated in this section.

\section{Conclusions}

The standard setup of two plates in relative motion, commonly used to study quantum friction, has been shown to be unstable in the nondissipative limit~\cite{brevik2022fluctuational,oue2024stable, Friction3}. Therefore, when calculating time-independent friction, it is essential to verify the system's stability within the relevant parameter range; otherwise, the calculation becomes invalid. In the nondissipative limit, the usual framework cannot be applied, as the system remains unstable even at arbitrarily small velocities 
$v$. 

In this work, we analyzed the friction force $F$
across various regimes, with a particular focus on behavior near the instability threshold, where the force diverges. Specifically, we demonstrated analytically that the transition into the unstable regime is accompanied by a logarithmic singularity in the friction force.

Let us briefly mention several additional problems not treated in this brief paper:

\noindent
(I)
Instead of a pair of plates, one often considers a nanoparticle moving parallel to a surface, which can also be unstable~\cite{silveirinha2014optical} and experience a divergent force.

\noindent
(II) 
What about the ``usual'' vertical Casimir-Lifshitz force $F _ z$?
Since the instability originates from the denominator in Eq.~\eqref{eq:F} (multiple scattering) and the expression for $F _ z$ (see, e.g., Ref.~\cite{dedkov2017fluctuation}) has exactly the same denominator, we expect that $F _ z$ should diverge when the instability threshold is approached.

\noindent
(III)
Dragging plates mechanically, with significant velocity and precision is extremely challenging.
Therefore, an alternative setup has been proposed in the literature (see Refs.~\cite{volokitin2011quantum,shapiro2017fluctuation,morgado2018drift} and references therein), namely: The plates are at rest, but a stationary drift current is induced in one of the plates.
Indeed, while the velocity $v \sim 10 ^ 5 \ \mathrm{m/s}$ that we took as an example when discussing \figref{fig:comparison} may seem too large if we consider a translational mechanical motion, it is not unrealistic in the context of electron drift velocities under high electric fields (saturation velocity)~\cite{quay2000temperature}.
There are many similarities (and some significant differences) between this ``electrical'' setup and the more standard, mechanical one, as discussed in detail in Ref.~\cite{shapiro2017fluctuation}.
There is a new twist in the ``electrical'' setup regarding a possible instability: It is well known that a semiconductor sample, subjected to a drift current, can become unstable (such instabilities have been thoroughly studied in the past, see Ref.~\cite{sydoruk2010terahertz} and references therein).
This means that even a single isolated plate with a drift current in it can be unstable.

These additional considerations underscore the complexity and richness of quantum friction phenomena, suggesting that the parallels between mechanical and electrical setups could provide further insights into fluctuation-induced forces in nanoscale systems near criticality.

\begin{acknowledgments}
This work has been supported by the Institution of Engineering and Technology (IET) and by FCT/MECI through national funds and when applicable co-funded EU funds under UID/50008: Instituto de Telecomunica\c{c}\~oes under project UIDB/50008/2020.
D.\,O.\,is supported by the JSPS Overseas Research Fellowship. 
\end{acknowledgments}



\bibliography{refs}

\end{document}